\def\year{2020}\relax
\documentclass[letterpaper]{article} 
\usepackage{aaai20}  
\usepackage{times}  
\usepackage{helvet} 
\usepackage{courier}  
\usepackage[hyphens]{url}  
\usepackage{graphicx} 
\usepackage{graphicx}  
\usepackage{mathrsfs}
\usepackage{mathtools}
\usepackage{amsmath}
\usepackage{arydshln}
\usepackage{multicol}
\usepackage{blkarray}
\usepackage{enumerate}
\usepackage{multirow}
\usepackage{rotating}
\usepackage{booktabs}
\usepackage{algorithm}
\usepackage{algorithmic}
\usepackage{cases}
\usepackage{subfigure}
\usepackage{amsfonts}
\usepackage{color}
\newtheorem{theorem}{Theorem}

\title{Discrete Knowledge Graph Embedding based on Discrete Optimization}

\author{Yunqi Li$^1$, Shuyuan Xu$^1$, Bo Liu$^2$, Zuohui Fu$^1$, Shuchang Liu$^1$, Xu Chen$^3$, Yongfeng Zhang$^1$\\
$^1$Department of Computer Science, Rutgers University, USA\\
$^2$JD Finance America Corporation, Mountain View, California, USA\\
$^3$Department of Computer Science, University College London, UK\\
\{yunqi.li, shuyuan.xu\}@rutgers.edu, kfliubo@gmail.com, \{zuohui.fu, shuchang.syt.liu\}@rutgers.edu,\\ successcx@gmail.com, yongfeng.zhang@rutgers.edu\\
}
\begin{document}

\maketitle

\begin{abstract}
This paper proposes a discrete knowledge graph (KG) embedding (DKGE) method, which projects KG entities and relations into the Hamming space based on a computationally tractable discrete optimization algorithm, to solve the formidable storage and computation cost challenges in traditional continuous graph embedding methods. The convergence of DKGE can be guaranteed theoretically. Extensive experiments demonstrate that DKGE achieves superior accuracy than classical hashing functions that map the effective continuous embeddings into discrete codes. Besides, DKGE reaches comparable accuracy with much lower computational complexity and storage compared to many continuous graph embedding methods.

\end{abstract}

\section{Introduction}
A knowledge graph (KG) is a multi-relational graph, whose nodes correspond to entities that represent objects or concepts, and edges represent different types of relations between the connected nodes. 
Graph embedding refers to learning a numerical feature representation for graph nodes and edges. 
Existing KG embedding techniques typically learn the node and edge feature in the continuous feature space, which usually imposes formidable challenges in storage and computation costs, particularly in large-scale applications. Take the widely used translation-based model TransE \cite{bordes2013translating} as an example, even as a shallow model, TransE requires about 33GB memory to store parameters when applied to Freebase \cite{bollacker2008freebase} with dimension of 200, which prohibits its application in limited resource scenarios such as edge devices. 
Therefore, it is necessary to learn more compact embeddings to solve the problems. Discrete code learning (hashing), which aims to map the original high-dimensional data into a low-dimensional discrete coding space with similarity preservation, has got great gains in storage and computation. Due to the difficulty of resolving discrete constraints, existing hashing methods usually adopt a two-stage learning process: relaxed optimization by discarding the discrete constraints first, and then binary quantization over the learned continuous embeddings. It has been argued that such a method will compromise the performance of large-scale problem and result in a large quantization loss \cite{zhang2016discrete}. 

In this paper, we propose a novel discrete knowledge graph embedding (DKGE) approach for learning more compact representations through discrete optimization directly. 
We use Hadamard product to represent the translation in Hamming space, and after translation, the Hamming distance between the connected entities in the graph is smaller 
than those unconnected entities. 
A computationally tractable discrete optimization algorithm is introduced to solve the learning objective through mixed-integer alternating optimization \cite{zhang2016discrete}. 
The convergence of our algorithm can be guaranteed theoretically. We conduct experiments on the FB15k-237 and WN18RR datasets for the link prediction task 
in KG, which show that DKGE achieves lower computational complexity and storage than state-of-the-art continuous KG embedding methods, while keeping comparable accuracy. It also achieves superior accuracy than directly hashing the effective continuous embeddings into discrete codes through three classical hashing functions. We also compare our method with another discrete knowledge  graph  embedding  method  named  B-CP \cite{kishimoto2019binarized} and show that DKGE gives much better accuracy on a more dense dataset.

\section{Related Work}
Knowledge graph models information in the form of entities and relationships between them. KG embedding is to embed entities and relations of a KG into vector spaces. Prominent examples include the Translational Embedding (TransE) model \cite{bordes2013translating} and its extensions such as TransH \cite{wang2014knowledge}, RESCAL \cite{nickel2011three}, etc. Learning to hash aims to learn data-dependent and task-specific hash functions, which can yield compact discrete codes to achieve efficiency \cite{liu2011hashing} and search accuracy \cite{wang2012semi}. \cite{kishimoto2019binarized} proposed a binary KG embedding method based on CANDECOMP/PARAFAC (CP) tensor decomposition, which binarizes continuous embeddings to discrete embeddings based on a quantization function in each learning step. In this paper, however, we propose to learn binary codes through solving discrete optimization directly, which gives better efficiency and accuracy on dense datasets.

\section{Discrete Knowledge Graph Embedding}
In this section, we present our Discrete Knowledge Graph Embedding (DKGE) method, which learns the discrete representations of entities and relations for KG based on discrete optimization. 
\subsection{Preliminaries}
Given a knowledge graph consisting of $n$ entities and $m$ relations, we can represent it as a set $S$ of triplets $\{(h,r,t)\}$, which is composed of a head entity $h\in\mathcal{E}$, a tail entity $t\in\mathcal{E}$, and a relation $r\in\mathcal{R}$ connecting them, where $\mathcal{E}$ is the set of all the entities and $\mathcal{R}$ is the collection of all the relations. We use bold uppercase and lowercase letters to represent matrices and vectors, respectively. In particular, we use \textbf{e} as the entity vector, including head entity vector \textbf{h} and tail entity vector \textbf{t}, and use \textbf{r} as the relation vector. $\textbf{E}=[\textbf{e}_1,\dots,\textbf{e}_n]$ represents the matrix of all the entities and $\textbf{R}=[\textbf{r}_1,\dots,\textbf{r}_m]$ is the matrix consisting of all the relations in KG. We denote tr($\cdot$) as the trace of a matrix, sgn($\cdot$): $\mathbb{R}\to \{\pm 1\}$ as the round-off function, and $\|\cdot\|_F$ represents the Frobenius norm of a matrix.

In DKGE, the embeddings $\textbf{h}$, $\textbf{t}$ and $\textbf{r}$ take discrete values as $\{\pm1\}^k$. $\textbf{E}=[\textbf{e}_1,\dots,\textbf{e}_n]\in \{\pm1\}^{k \times n}$ and $\textbf{R}=[\textbf{r}_1,\dots,\textbf{r}_m]\in \{\pm1\}^{k \times m}$ are the matrices of $k$-length entity and relation discrete codes, respectively. 
Here, we propose to use the Hadamard product of \textbf{h} and \textbf{r}, which still maps the discrete codes \textbf{h} and \textbf{r} into the Hamming space, to represent the translation of the head entity under this relation. Intuitively, the translation from the head embedding \textbf{h} to its tail embedding \textbf{t} connected by relation \textbf{r} imitates ``jumping'' along the axis in the direction of \textbf{r} indicates. 

The Hamming distance between \textbf{t} and the Hadamard product of \textbf{h} and \textbf{r} is defined as:
\begin{equation}
\begin{aligned}
&d(\textbf{h} \times \textbf{r},\textbf{t})
=\sum_{i=1}^{k} \mathbb{I} ((\textbf{h} \times \textbf{r} )_i \neq  \textbf{t}_i)\\
= &\frac{1}{2} (k+\sum_{i=1}^{k} (\mathbb{I} ((\textbf{h} \times \textbf{r} )_i \neq  \textbf{t}_i) -\mathbb{I} ((\textbf{h} \times \textbf{r} )_i=  \textbf{t}_i)))\\
=&\frac{1}{2} (k-\sum_{i=1}^{k}((\textbf{h} \times \textbf{r} )_i^{\mathrm{T}}\textbf{t}_i)=\frac{1}{2}(k-(\textbf{h} \times \textbf{r})^{\mathrm{T}} \textbf{t})
\end{aligned}
\end{equation}
where $\mathbb{I}(\cdot)$ denotes the indicator function that returns 1 if the statement is true and 0 otherwise. 
We also design the scoring function of DKGE to imply the Hamming distance between $\textbf{h}\times\textbf{r}$ and \textbf{t}. Given a fact $(h,r,t)$, \textbf{t} should be the nearest neighbor of $\textbf{h}\times\textbf{r}$. Therefore, the problem of DKGE is formulated as:

\begin{equation}
\begin{aligned}
\begin{split}
	\mathop{\arg\min}_{\textbf{E},\textbf{R}} \ \ &\sum\limits_{(h,r,t)\in S} \sum\limits_{(h^\prime,r,t^\prime)\in S^\prime} [\gamma-[(\textbf{h} \times \textbf{r})^{\mathrm{T}} \textbf{t}-(\textbf{h}' \times \textbf{r})^{\mathrm{T}} \textbf{t}']]_+ \\
	&s.t.~~~\textbf{E}\in \{\pm1\}^{k \times n}, \textbf{R}\in \{\pm1\}^{k \times m};\\
	&\underbrace{\textbf{E}\textbf{1}=0,\textbf{R}\textbf{1}=0}_{\textup{Balanced Partition}}, \  
	\underbrace{\textbf{E}\textbf{E}^\mathrm{T}=n\textbf{I}, \textbf{R}\textbf{R}^\mathrm{T}=m\textbf{I}}_{\textup{Decorrelation}},
\end{split}
\end{aligned}
\end{equation}
where $\gamma$ is a margin hyper-parameter, $[x]_+$ denotes the hinge loss, $S'_{(h,r,t)}=\{(h',r,t) |h' \in\mathcal{E}\}\bigcup\{(h,r,t') |t'\in\mathcal{E}\}$ is the set of corrupted triplets, where either the head or the tail entity of each training triplet is replaced by a random entity. 
Since the regularizers $\left \|\textbf{E}\right \|_F^2$ and $\left \|\textbf{R}\right \|_F^2$
are constants given a fixed dimension under discrete codes, so we do not need to consider them here. Besides, to hash the entities and relations in a more informative and compact way, we need to impose two additional constraints called Balanced Partition and Decorrelation \cite{weiss2009spectral} to maximize the information entropy of the bits, which requires each bit to split the dataset as balanced and uncorrelated as possible to remove the redundancy among the bits. 

\subsection{Learning Model}
Actually, it is challenging to solve DKGE in Eq.(2), since it is in general an NP-hard problem \cite{haastad2001some}. In this subsection, we introduce a learning model that solves DKGE in a computationally tractable manner by softening the balance partition and decorrelation constraints \cite{zhang2016discrete}. Denote $\mathcal{X}=\{\textbf{X}\in\mathbb{R}^{k\times n}|\textbf{X}\textbf{1}=0,\ \textbf{X}\textbf{X}^\mathrm{T}=n\textbf{I}\}$,  $\mathcal{Y}=\{\textbf{Y}\in\mathbb{R}^{k\times m}|\textbf{Y}\textbf{1}=0,\ \textbf{Y}\textbf{Y}^\mathrm{T}=m\textbf{I}\}$, as well as distances $d(\textbf{E},\mathcal{X})=\mathop{\min}_{\textbf{X}\in\mathcal{X}}\|\textbf{E}-\textbf{X}\|_F $ and $d(\textbf{R},\mathcal{Y})=\mathop{\min}_{\textbf{Y}\in\mathcal{Y}}\|\textbf{R}-\textbf{Y}\|_F $, Eq.(2) can be relaxed as:
\begin{equation}
\begin{aligned}
\begin{split}
	\mathop{\arg\min}_{\textbf{E},\textbf{R},\textbf{X},\textbf{Y}} \ \ &\sum\limits_{(h,r,t)\in S} \sum\limits_{(h^\prime,r,t^\prime)\in S^\prime} [\gamma-[(\textbf{h} \times \textbf{r})^{\mathrm{T}} \textbf{t}-(\textbf{h}' \times \textbf{r})^{\mathrm{T}} \textbf{t}']]_+ \\
	&+ \alpha d^2(\textbf{E},\mathcal{X})+ \beta d^2(\textbf{R},\mathcal{Y})\\
	&s.t.~~~\textbf{E}\in \{\pm1\}^{k \times n}, \textbf{R}\in \{\pm1\}^{k \times m},
\end{split}
\end{aligned}
\end{equation}
where $\alpha>0$ and $\beta>0$ are tuning parameters. Actually, we can further simplify the above Eq.(3) as the following form:
\begin{equation}
\begin{aligned}
\begin{split}
	\mathop{\arg\min}_{\textbf{E},\textbf{R},\textbf{X},\textbf{Y}} \ \ &\sum\limits_{(h,r,t)\in S} \sum\limits_{(h^\prime,r,t^\prime)\in S^\prime} [\gamma-[(\textbf{h} \times \textbf{r})^{\mathrm{T}} \textbf{t}-(\textbf{h}' \times \textbf{r})^{\mathrm{T}} \textbf{t}']]_+ \\
	&-2\alpha \text{tr}(\textbf{E}^\mathrm{T}\textbf{X})-2 \beta \text{tr}(\textbf{R}^\mathrm{T}\textbf{Y})\\
	&s.t.~~~\textbf{E}\in \{\pm1\}^{k \times n}, \textbf{R}\in \{\pm1\}^{k \times m},\\
	&\textbf{X}\textbf{1}=0, \textbf{X}\textbf{X}^\mathrm{T}=n\textbf{I};\ \textbf{Y}\textbf{1}=0, \textbf{Y}\textbf{Y}^\mathrm{T}=m\textbf{I}
\end{split}
\end{aligned}\label{eq:loss}
\end{equation}

Now we have Eq.(4) as our learning model for DKGE, which allows a certain discrepancy between the discrete codes and the corresponding continuous values. Therefore, it is formalized as a joint mix-integer optimization problem, which can be solved by the four subproblems in \cite{zhang2016discrete}. Moreover, the convergence of DKGE can also be guaranteed theoretically in a similar way.

\section{Experiments}

\begin{table*}[t]
    \centering
    \begin{tabular}{|l|ccccc|ccccc|}
    \hline
        Dataset & \multicolumn{5}{c|}{WN18RR} & \multicolumn{5}{c|}{FB15k-237}\\
        Evaluation & MR & MRR & Hit@1 & Hit@3 & Hit@10 & MR & MRR & Hit@1 & Hit@3 & Hit@10\\\hline
        DistMult* & 7000 & 0.444 & 41.2 & \textbf{47.0} & 50.4 & 512 & 0.281 & 19.9 & 30.1 & 44.6\\
        ComplEx* & 7882 & 0.449 & 40.9 & 46.9 & 53.0 & 546 & 0.278 & 19.4 & 29.7& 45.0\\
        TransE* & 2300 & 0.243 & 4.3 & 44.1 & 53.2 & 323 & 0.279 & 19.8 & 37.6 & 44.1\\
        ConvE* & 4464 & \textbf{0.456} & 41.9 & \textbf{47.0} & 53.1 & 245 & 0.312 & 22.5 & 34.1 & 49.7\\
        ConvKB* & \textbf{1295} & 0.265 & 5.8 & 44.5 & \textbf{55.8} & \textbf{216} & 0.289 & 19.8 & 32.4 & 47.1\\
        R-GCN* & 6700 & 0.123 & 8.0 & 13.7 & 20.7 & 600 & 0.164 & 10.0 & 18.1 & 30.0\\\hline
        TransE+sign & 2635 & 0.129 & 0.9 & 19.2 & 35.3 & 429 & 0.214 & 14.6 & 23.6 & 33.9\\
        TransE+equal & 12315 & 0.073 & 0.9 & 10.7 & 18.3 & 6883 & 0.024 & 2.1 & 2.4 & 2.7\\
        TransE+Lloyd & 8649 & 0.100 & 0.6 & 15.5 & 27.0 & 5716 & 0.011 & 0.6 & 1.1 & 1.7\\
        DistMult+sign & 8513 & 0.023 & 1.1 & 2.4 & 4.1 & 3021 & 0.015 & 0.7 & 1.2 & 2.5\\
        DistMult+equal & 27787 & 0.000 & 0.0 & 0.0 & 0.0 & 6229 & 0.004 & 0.1 & 0.3 & 0.6\\
        DistMult+Lloyd & 15627 & 0.001 & 0.0 & 0.0 & 0.2 & 6231 & 0.006 & 0.4 & 0.5 & 0.7 \\\hline
        B-CP(D=128) & - & 0.413 & 38.5 & 43.1 & 46.1 & - & 0.214 & 15.2 & 22.7 & 33.7\\
        B-CP(D=256) & - & 0.453 & \textbf{43.5} & 46.0 & 48.7 & - & 0.268 & 18.8 & 29.2 & 42.7\\
        B-CP(D=512) & - & 0.444 & 42.5 & 45.2 & 48.0 & - & 0.284 & 19.3 & 31.4 & 46.7\\
        \hline
        \textbf{DKGE}(D=128) & 5217 & 0.369 & 31.6 & 40.1 & 46.8 & 551 & 0.342 & 28.9 & 36.6 & 43.3\\
        \textbf{DKGE}(D=256) & 5225 & 0.392 & 34.0 & 42.4 & 49.5 & 464 & \textbf{0.416} & \textbf{36.8}& \textbf{43.2} & \textbf{50.7}\\
        \textbf{DKGE}(D=512) & 5469 & 0.403 & 35.0 & 43.3 & 50.6 & 474 & 0.396 & 34.5 & 41.4 & 48.7\\\hline
        
    \end{tabular}
    \vspace{-8pt}
    \caption{The results for link prediction task. * indicates the results are transcribed from \cite{KBGAT2019}. MR is the lower the better, while MRR and Hits are the higher the better. Sign, equal, and Lloyd are three quantization methods \cite{lloyd1982least}. The best results are in \textbf{bold}. The dimension of TransE and DistMult continuous embeddings are 64, meanwhile the quantizer use 8 bits to represent one continuous value.}
    \label{tab:predictionresult}
\vspace{-10pt}
\end{table*}
\subsection{Experiment Setting}
To show the feasibility of our discrete model, we test our model on two benchmark datasets: WN18RR and FB15k-237. 
For evaluation, we test our approach on the link prediction task, which aims to predict a triplet with head or tail entity missing. Specifically, we replace the head or tail entity with every other entity to compute the scores for each triplet and sort the scores to record the correct rank. Similarly to previous work \cite{nguyen2017novel,dettmers2018convolutional}, to get more reliable results, we report the performance on the \textit{filtered} setting, i.e, remove the triplets that present in the training and validation set during the evaluation. We report mean rank (MR), mean reciprocal rank (MRR) and the hit rate in the top $N$ for $N=1,3,10$, which are classical evaluating indicators used in KG embedding models. We test our model with embedding dimension \{32,64,128,256,512\}, $\gamma$ is tuned in \{32,64,128,256,512,1024\}, and the trade-off parameters $\alpha,\beta$ are tuned in \{1e-7,1e-5,1e-3,0.1,1\}. We report the performance with the best setting.

\begin{figure*}[t]
\mbox{
\centering
    \subfigure[WN18RR]{\label{fig:WN18RR_mrr}
	    \includegraphics[width=0.25\textwidth]{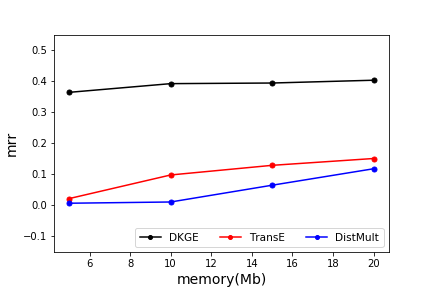}}
	    \hspace{-15pt}
	\subfigure[FB15k-237]{\label{fig:FB15k237_mrr}
	    \includegraphics[width=0.25\textwidth]{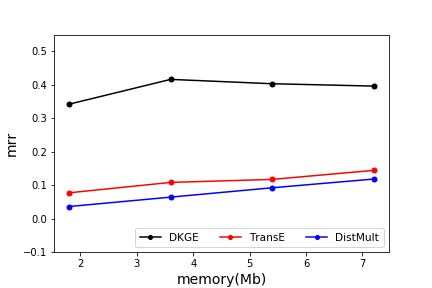}}
	    \hspace{-15pt}
	\subfigure[WN18RR]{\label{fig:WN18RR_hit10}
	    \includegraphics[width=0.25\textwidth]{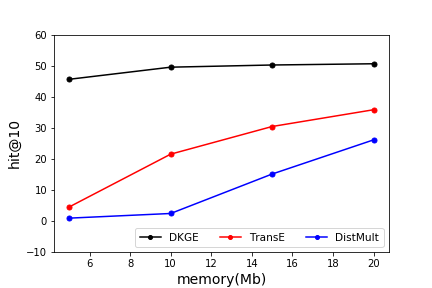}}
	    \hspace{-15pt}
	\subfigure[FB15k-237]{\label{fig:FB15k237_hit10}
	    \includegraphics[width=0.25\textwidth]{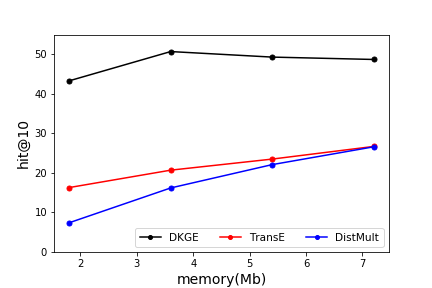}}
}
\vspace{-12pt}
\caption{The performance within same memory. The memory is just used to store the embedding value of entities and relations.}
\label{fig:memory}
\vspace{-10pt}
\end{figure*}

\subsection{Results and Discussions}

\textbf{DKGE vs. Continuous Models:} 
In this part, we compare the performance of DKGE with some continuous methods, including classical approaches such as TransE \cite{bordes2013translating}, DistMult \cite{yang2014embedding} and ComplEx \cite{trouillon2016complex}, as well as some state-of-the-art methods such as ConvE \cite{dettmers2018convolutional}, 
ConvKB \cite{nguyen2017novel} and R-GCN \cite{schlichtkrull2018modeling}. We report the results in Table \ref{tab:predictionresult}. We can see that our DKGE model can achieve comparable performance against continuous approaches with much lower computational complexity and storage on both of the two datasets, and it is better on at least one measure. 
To illustrate the memory saving of our DKGE model more rigorously, we compare the performance of our approach with continuous models within the same memory cost, as shown as Fig \ref{fig:memory}. Besides, by representing the entities and relations as discrete codes, we can conduct faster computation with bitwise operations among discrete vector representations, which greatly reduces the running time of DKGE required to achieve various tasks on KG.

\textbf{DKGE vs. Hashing functions:} In this part, we use three classical hashing functions to directly hash the continuous embeddings obtained by TransE and Distmult into discrete codes, and compare the performance of those embeddings with our learned DKGE embedding. The first hashing method is discretizing the continuous embedding with sign function; the second one is discretizing the representations by equally divided intervals, specifically, we divide the range of the embedding values into $n$ intervals evenly, and use a binary vector of length $m$ to represent each interval. Then we can discretize the $d$-dimension continuous embedding into a $(m*d)$-dimension binary embedding; the third one is using Lloyd-Max Algorithm \cite{lloyd1982least}, which is a typical quantization algorithm commonly used in data compression and digitizing continuous-valued signals. The key difference of the last two approaches is that we divide the range of the embedding values into $n$ intervals according to a more complex distribution but not evenly. Overall, we can see that our discrete optimization model obtains much better performance than discretizing the continuous embedding values, which shows the advantage of our discrete learning approach against traditional quantization algorithms based on continuous-valued embeddings.

\textbf{DKGE vs. B-CP:} Different from the above methods that conducts quantization after the whole continous optimization procedure has concluded, B-CP \cite{kishimoto2019binarized} conducts quantization during each iteration of the continuous optimization procedure. Our DKGE approach, however, moves one step further by conducting discrete optimization directly. Besides, we use only one vector to represent each entity and relation, while B-CP learns two vectors for each entity (for head and tail, respectively), and two vectors for each relation (relation and its inverse vector). As a result, our memory consumption is only half of B-CP under the same embedding dimension. As for the performance, we can see that our model performs much better than B-CP on FB15k-237 with only half of their memory space. On WN18RR, DKGE gets better Hit@10, the less favorable performance of DKGE on other metrices implies that discrete optimization tends to perform better on dense datasets, where each relation only needs to translate a small group of entities for better discrimination in the discrete embedding space.

\section{Conclusions}
This paper focuses on the challenging problem of seeking compact representations for knowledge graphs by discrete optimization. We propose a novel hashing approach called discrete knowledge graph embedding (DKGE) to embed entities and relations of KG into discrete codes, which helps to solve the formidable challenges of storage and computation cost in continuous embeddings. We design a Hamming distance-based scoring function, and develop an efficient algorithm to learn the discrete embeddings directly through discrete optimization, which can be solved in a computationally tractable manner by an alternating optimization algorithm. Moreover, the convergence of the proposed method can be theoretically guaranteed. We operate a series of experiments to show that DKGE is able to achieve satisfactory accuracy while enjoying both computational and memory efficiency than those classical and even state-of-the-art continuous knowledge graph embedding methods. 
In the future, we will explore other translational operations beyond Hadamard product for discrete embedding, as well as disentangled embeddings in the discrete space for intepratable knowledge graph embedding.

\small
\bibliography{aaai_2020}
\bibliographystyle{aaai}

\end{document}